\documentclass[aps,twocolumn,nofootinbib,superscriptaddress,prd]{revtex4-2}
\usepackage{graphicx}
\usepackage{amsmath}
\usepackage{slashed}
\usepackage{amssymb}
\usepackage{amsfonts}
\usepackage{mathtools}
\usepackage{color}
\usepackage{dsfont}
\usepackage{soul}
\usepackage{hyperref}
\usepackage{tikz}
\usetikzlibrary{positioning,shapes}
\DeclareMathOperator{\sinc}{sinc}
\DeclareMathOperator{\arsinh}{arsinh}
\DeclareMathOperator{\sech}{sech}
\usepackage{tikz,xcolor}
\definecolor{lime}{HTML}{A6CE39}
\DeclareRobustCommand{\orcidicon}{%
	\begin{tikzpicture}
	\draw[lime, fill=lime] (0,0) 
	circle [radius=0.16] 
	node[white] {{\fontfamily{qag}\selectfont \tiny ID}};
	\draw[white, fill=white] (-0.0625,0.095) 
	circle [radius=0.007];
	\end{tikzpicture}
	\hspace{-2mm}
}
\foreach \x in {A, ..., Z}{%
	\expandafter\xdef\csname orcid\x\endcsname{\noexpand\href{https://orcid.org/\csname orcidauthor\x\endcsname}{\noexpand\orcidicon}}
}

\begin{document}
\title{Quantum vibrational mode in a cavity confining a massless spinor field}
\date{\today}

\author{Alessandro Ferreri\orcidA{}}
\affiliation{Institute for Quantum Computing Analytics (PGI-12), Forschungszentrum J\"ulich, 52425 J\"ulich, Germany}

\begin{abstract}
We analyse the reaction of a massless (1+1)-dimensional spinor field to the harmonic motion of one cavity wall. In our model, the oscillation amplitude of the harmonic oscillator is promoted to a quantum operator, providing the system with an additional quantum degree of freedom having bosonic nature. After obtaining the interaction Hamiltonian, we estimate the correction to both the ground state and its energy. We demonstrate that the system is able to convert bosons into fermion pairs at the lowest perturbative order. Extension of our model to multiple bags is contemplated. 
\end{abstract}

\maketitle


\section{Introduction}
Despite their intrinsically different nature, fermionic and bosonic quantum fields share similar features when found in their ground state. As an example, in both cases the standard quantization procedure leads to the presence of a residual vacuum energy, also called Casimir energy, which plays an active role when fields are confined in a finite region of space \cite{plunien_casimir_1986}.
Generally, the mathematical formalism to confine quantum fields is carried out differently for bosonic and fermionic systems. Whilst the confinement of bosonic fields typically occurs by imposing specific boundary conditions on the equations of motion \cite{bordag_new_2001}, for spinor fields the boundary conditions are included in the Hamiltonian of the system, and are brought about by a well potential consisting of sharp walls, also called Dirac spikes \cite{sundberg_casimir_2004}. It is relevant to notice that the presence of the above-mentioned vacuum energy induces an attractive force between the cavity walls, as expected by the Casimir effect \cite{sundberg_casimir_2004,  fosco_functional_2008, zhabinskaya_casimir_2008}.

Beyond the static scenario, further interesting phenomena emerge when the cavity undergoes some non-inertial motion, the latest described in terms of time-dependent boundary conditions for the confined field \cite{friis_scalar_2013}. One of the most fascinating effects ascribable to dynamical systems is the creation of particles, which occurs whenever the vacuum state of the quantum field under motion does not correspond to any vacuum state of the same field in an inertial frame of reference \cite{birrell1984quantum}. 
As an example, let us assume that one of the cavity walls confining the field undergoes an oscillating motion. In this case, the modulation of the boundary condition alters the modes of the field, and as a consequence, an inertial detector would reveal particles even if the field was initially prepared in its vacuum state.
Such phenomenon, called dynamical Casimir effect (DCE), was initially predicted for the electromagnetic field \cite{doi:10.1063/1.1665432, dodonov_fifty_2020}, and afterwards extended to spinor fields \cite{diego_mazzitelli_fermions_1987, fosco_quantum_2007, fosco_dynamical_2022}.

In the first formulation of the DCE, it was assumed that the motion of the cavity wall is strictly determined by the time-dependence of the boundary conditions the field is subject to. However, in a generic scenario, one can suppose that the wall neither undergoes a motion imposed by an external force, nor is fixed, but it is found unfastened at some position. In this case, it can be thought that such wall (and consequently the length of the cavity) undergoes some fluctuation, which can be mathematically described with the same formalism of a quantum harmonic oscillator \cite{law_interaction_1995,  ferreri2022interplay}.
This idea, on which quantum optomechanics is based \cite{aspelmeyer_cavity_2014, meystre_short_2013}, suggests that the position of the wall can be treated as an additional quantum mechanical degree of freedom of the system, and that the particle creation as predicted by the DCE can be thought as an excitation transfer between the cavity field and the additional mechanical mode \cite{ferreri2022interplay, macri_nonperturbative_2018}. 

In this manuscript, we want to apply these concepts to a system consisting of a spinor field confined in a bag by two Dirac spikes. In particular, following the procedure developed in \cite{ferreri2022interplay}, we are going to induce a fluctuation in the position one of the two spikes, and promote such position to a quantum operator acting on a vibrational degree of freedom. The Hamiltonian of the whole system resulting from this formalism accounts for both the optomechanical coupling and the excitation exchange between field and oscillating wall; moreover, it includes counterrotating terms which give rise to a shift of the energy of the ground state. In our model, the fermionic particle creation can arise from both the conversion of bosonic excitation and from the mechanical stimulation of a time-dependent external drive acting on the vibrational mode. 

The article is structured as follows: in Sec.\ref{tm} we present the mathematical procedure to introduce the quantum vibrational mode to a system consisting of a spinor field confined between two infinite spikes. In Sec.\ref{corr} we evaluate the corrections to both ground state and its energy caused by the interaction between fermionic and bosonic degrees of freedom. In Sec.\ref{TP} we present the transition probability to generate fermionic particles starting from a generic input bosonic state and under the action of the external drive. We calculate such probability for some concrete scenarios, using different bosonic states, as well as an impulsive drive. In Sec.\ref{extension} we extend our model to a scenario wherein multiple spikes are unfastened, hence free to vibrate. Finally, conclusions are given in Sec.\ref{concl}.
\section{Theoretical model}\label{tm}
In this section we present the procedure to obtain the Hamiltonian of a confined spinor field undergoing harmonic boundary conditions. This
is accomplished by adapting the formalism developed in \cite{ferreri2022interplay} to the fermionic case.
\subsection{Hamiltonian of the system}\label{Hamilton}
We begin our analysis by considering a spinor field subject to a static potential in 1+1 dimension.
Adopting the natural units, $\hbar=c=1$, the Lagrangian of such field is:
\begin{align}
\mathcal{L}(x,t)=\bar{\psi}(x,t)(i\slashed{\partial}-M-V(x))\psi(x,t),
\end{align}
where $M$ is the mass, the adjoint field is defined as $\bar\psi=\psi^\dag\gamma^0$, and the potential,
\begin{align}
V(x)=g_L\delta(x)+g_R\delta(x-L),
\label{pot}
\end{align}
determines the static (bag) boundary conditions of the spinor field confined in the region of space $x\in[0,L]$. The coupling constants $g_L$ and $g_R$, where $L$ and $R$ stand for ``left" and ``right" respectively, quantify the degree of confinement of the field, and it was shown that boundary conditions without imperfections are fulfilled whenever $g_L=g_R=2$ \cite{fosco_functional_2008}.

By solving the Dirac equation, we can express the spinor field as:
\begin{align}
\psi(x,t)=\sum_n\left(e^{-i\omega_n t}\psi_{n+}(x) c_n+e^{i\omega_n t}\psi_{n-}(x) d_n^*\right)
\end{align}
where $c_n$ and $d_n$ are the amplitudes for the fermionic particle and antiparticle respectively, and 
\begin{align}
\psi_{n\pm}(x)=\frac{1}{\sqrt{L}}
\begin{pmatrix}
\pm\sin(k_n z)\\
\cos(k_n z)
\end{pmatrix}.
\end{align}

An analytical dispersion relation for a generic confined spinor field in 1+1 dimension does not exist, but discrete analytical solutions can be found in the nonrelativistic ($M L\gg 1$) and in the massless ($M=0$) limits \cite{mamaev_vacuum_1980, fosco_dynamical_2022}. In this paper, we focus our investigation on the latter case, therefore the dispersion relation reads $\omega_n=k_n=\left(n+\frac{1}{2}\right)\frac{\pi}{L}$.

So far, the system was supposed classical and static. In order to account for the dynamics of the environment, we adapt the protocol developed in \cite{ferreri2022interplay} for a scalar bosonic field to the fermionic scenario under consideration. As a starting point, we assume that the right wall of the potential well undergoes a fluctuation of the form $L\rightarrow L+\delta L$, where $\delta L$ is the increment. Supposing that such increment is extremely small, we can Taylor-expand the delta function on the right side up to the first perturbation order, $\delta(x-L-\delta L)\simeq -\delta L\,\delta'(x-L)$, where $\delta'(x-x_0)$ indicates the derivative of the delta function located around the point $x_0$. As usual, the classical Hamiltonian is obtained by means of $H(x,t)=\int d^3x\left\{\frac{\partial \mathcal{L}}{\partial \dot\psi} \dot\psi+\frac{\partial \mathcal{L}}{\partial \dot{\bar\psi}} \dot{\bar\psi}-\mathcal{L}\right\}$. Finally, we promote the amplitudes of the spinor fields $c_n$ and $d_n$ (and their complex conjugates), as well as the increment $\delta L$, to quantum operators. The fermionic amplitudes turn to the fermionic annihilation and creation operators for the particle and the antiparticle respectively, $c_n\rightarrow \hat c_n$ and $d_n^*\rightarrow \hat d_n^\dag$. Such operators follows the standard anticommutation rules: $\{\hat{c_n},\hat{c_m}^\dag\}=\{\hat{d_n},\hat{d_m}^\dag\}=\delta_{nm}$, whereas all others vanish. The notation $\{\cdot,\cdot\}$  specifies the anticommutator.
Similarly, the increment $\delta L$ becomes proportional to the position operator $\hat X$ acting on an additional bosonic degree of freedom, $\delta L\rightarrow \delta L_0(\hat{b}^\dag+\hat{b})$, where $\delta L_0$ is the oscillation amplitude, and $\hat{b}^\dag$ and $\hat{b}$ are, respectively, the creation and annihilation operators of a quantum harmonic oscillator having frequency $\Omega$ and representing the vibrational degree of freedom of the right wall. Such operators fulfill the standard bosonic commutation rule $[\hat b,\hat b^\dag]=1$, where the notation $[\cdot,\cdot]$ specifies the commutator. 

By following the procedure reported above, we are in the position to write the Hamiltonian operator of the whole system, where we can distinguish the quantum spinor field, the quantum harmonic oscillator, as well as their interaction. In addition, we can apply external drives on both the fermionic and bosonic degree of freedoms. Assuming that the drive acts only on the oscillating wall, and that all operators are normal ordered, the total Hamiltonian reads: 
\begin{align}
\hat H(t)=&\hat H_0+\epsilon H_I+\hat H_{\textrm{dr}}(t)\label{htot},\\
\hat H_0=&\sum_n\omega_n(\hat c_n^\dag\hat c_n+\hat d_n^\dag\hat d_n)+\Omega\hat b^\dag\hat b,\\
\hat H_I=&-2\sum_{n m}(-1)^{n+m}\left[(\omega_n+\omega_m)(\hat c_n^\dag \hat c_m+\hat d_m^\dag\hat d_n)\right.\nonumber\\
&\left.+(\omega_n-\omega_m)(\hat d_n\hat c_m+\hat d_m^\dag\hat c_n^\dag)\right](\hat b^\dag+\hat b),\label{HI}\\
\hat H_{\textrm{dr}}(t)=&\lambda(t)\hat b^\dag+\lambda^*(t)\hat b, 
\end{align}
where we introduced the adimensional amplitude $\epsilon=\delta L_0/L$, and the complex drive is expressed as $\lambda(t)=\lambda_x(t)+i\lambda_p(t)$.

The generic eigenstate of the unperturbed Hamiltonian $\hat H_0$ is $\lvert x; y_1, y_2,...,y_n,...,\bar y_{1},\bar y_{2},...,\bar y_{m},...\rangle$, where $x\in\mathbb{N}$ indicates the eigenvalue of the bosonic subsystem, whereas $y_n$ and $\bar y_{m}$ indicate the presence of a fermionic particle in the mode $n$ and the antiparticle in the mode $m$. Hence, $y_n$ and $\bar y_{m}$ can assume either the value 0 or 1.

It is relevant to stress that the form of $\hat H_I$ is a direct consequence of our quantization procedure \cite{ferreri2022interplay}. Interestingly, such Hamiltonian operator contains terms expressing the optomechanical interaction between the field and harmonic oscillator, namely $\hat c_n^\dag \hat c_n(\hat b^\dag+\hat b)$ and $\hat d_n^\dag \hat d_n(\hat b^\dag+\hat b)$. Moreover, besides the counterrotating terms, the last line in Eq.\eqref{HI} expresses the possibility to convert single phonons into fermion pairs. In a similar manner, this phenomenon occurs in fully bosonic systems, where single phonons can be down-converted in photon pairs \cite{ferreri2022interplay, macri_nonperturbative_2018, settineri_conversion_2019}. However, in contrast to the bosonic case, we stress that the interaction Hamiltonian in Eq.\eqref{HI} cannot be linearized following the arguments in \cite{aspelmeyer_cavity_2014, meystre_short_2013}, since no coherent state can be defined for fermions. 

\section{Ground state and effective vacuum energy}\label{corr}
The structure of Eq.\eqref{HI} highlights that the interaction between field and wall can occur in different ways.
As a first example, we already noticed the presence of the optomechanical coupling, proportional to $(\hat c_n^\dag \hat c_m+\hat d_m^\dag\hat d_n)(\hat b^\dag+\hat b)$. This term describes the shift of the wall caused by the presence of quantum excitations of the field within the cavity. In fully bosonic systems, the translation of the wall is associated to the radiation pressure due to the presence  of a large amount of photons, typically preparing a coherent state in a single mode \cite{aspelmeyer_cavity_2014, meystre_short_2013}. On the contrary, fermionic modes cannot contain more than one particle, and the only way for the optomechanical coupling to play a relevant role in the dynamics of our (non-degenerate) system would be to excite different fermionic modes. We leave the investigation of the radiation pressure in fermionic optomechanical systems to a future work. 

Since the optomechanical coupling is proportional to the fermion number, we expect that its presence will alter both the eigenstates and eigenenergy of the unperturbed Hamiltonian, but it will not affect the vacuum state. However, the interaction Hamiltonian also contains terms altering the total amount of excitations, and in principle such terms can bring about modifications to the ground state of the total system. More specifically, we observe that such modifications are caused by counterrotating terms.
\subsection{Correction to the vacuum state}
When introducing the total Hamiltonian, we defined the eigenstates of the unperturbed Hamiltonian $\hat H_0$. However, even switching off the external drive, such states does not represent an eigenbasis for the total Hamiltonian because of the presence of the interaction Hamiltonian, Eq.\eqref{HI}. Nevertheless, since the strength of the interaction is proportional to the amplitude $\epsilon$, which is supposed to be relatively small, we expect that the actual eigenstates of the whole Hamiltonian do not differ substantially from the unperturbed states. Therefore we can think about retaining the unperturbed eigenstates of $\hat H_0$ and calculating the corrections due to the presence of the interaction Hamiltonian. In particular, we will focus our analysis on the correction to the unperturbed vacuum state, which is expected to not coincide to the ground state of the system anymore.

In order to estimate the correction to the vacuum state, we make use of the first order perturbation theory.
The correction to the state at the lowest significant order yields:
\begin{align}
\lvert\Psi_{0}\rangle=\lvert 0; 0,\bar 0\rangle+2\epsilon\sum_{nm}\frac{(-1)^{n+m}(\omega_n-\omega_m)}{\omega_n+\omega_m+\Omega}\lvert 1; 1_n,\bar 1_{m}\rangle.
\end{align}
The state can be normalized as $\lvert\tilde\Psi_{0}\rangle=Z^{-1/2}\lvert\Psi_{0}\rangle$, with normalization constant $Z=\langle\Psi_{0}\mid\Psi_{0}\rangle$.
We observe that the interaction between the field and the dynamical wall of the bag brings about a change to the ground state of the system; in particular, the ground state consists now of two parts: the first is the vacuum state of the system, whereas the second is the correction caused by the presence of counterrotating terms in the interaction Hamiltonian. Such terms do not fulfill the energy conservation, hence they cannot be responsible for the creation of real particles. It is relevant to notice that the new dressed-like ground state indicates the presence of entanglement between fermionic and bosonic channels \cite{butera_field_2013}.
\subsection{Energy shift}
Another consequence of the field-wall interaction in Eq.\eqref{HI} is the shift of the energy levels. Again, the small amplitude $\epsilon$ enables a perturbative approach; in particular, the energy correction to the ground state is estimated by means of second order perturbation theory, which yields:
\begin{align}
    \Delta E=-\sum_{nm}\frac{4\epsilon^2(\omega_n-\omega_m)^2}{\omega_n+\omega_m+\Omega}.
    \label{deltaE}
\end{align}
It should not come as a surprise that, in contrast with the bosonic scenario \cite{butera_field_2013, armata_vacuum_2015}, such energy correction excludes diagonal terms, in accordance with the Pauli exclusion principle.
The negative sign of the energy shift indicates that the interaction between the fermionic and bosonic subsystems leads to the lowering of the Casimir energy within the cavity  \cite{butera_field_2013}. Generally, such energy can be normalized by means of a cut-off frequency in the ultraviolet limit \cite{milonni2013quantum}, although different techniques are usually employed for the estimation of the Casimir force in scenarios involving fermions \cite{sundberg_casimir_2004, fosco_functional_2008, zhabinskaya_casimir_2008}.

We can summarize the results of this section as follows: the presence of the interaction Hamiltonian leads to the modification of the ground state, which contains virtual particles in the basis of the unperturbed Hamiltonian. Such virtual particles mediate the interaction between the field and the wall, and reduce the ground energy of the system.
\section{Boson-fermion excitation transfer}\label{TP}
The last line of Eq.\eqref{HI} contains four coupling terms. Two of them, i.e. the counterrotating terms, are responsible for the shift of the energy levels, as seen in the previous section. The other two, namely those proportional to $\hat d_n\hat c_m\hat b^\dag$ and $\hat d_m^\dag\hat c_n^\dag\hat b$, can fulfill the energy conservation, and contribute to the fermion-boson excitation transfer.
In order to analyse this phenomenon, in this section we calculate the transition probability to generate fermion pairs from the fermionic vacuum state. We specify that, due to the presence of the external drive, our model can predict the fermion generation by either the stimulation of the external drive acting on the bosonic mode, or by an effective excitation transfer between bosonic and fermionic channels. 
\subsection{Transition probability}
We want to estimate the transition probability to find a fermionic particle in the mode $k$ and an antiparticle in the mode $k'$ at time $t$.
If we assume that phonons are initially prepared in a pure state $\lvert\psi_i\rangle$, the transition probability becomes:
\begin{align}
P(t)=\lvert\langle\psi_f; 1_k,\bar 1_{k'}\lvert \hat U(t)  \rvert\psi_i; 0_n,\bar 0_{n}\rangle\rvert^2
\end{align}
where $\lvert\psi_f\rangle$ is the final phononic state and $\hat U(t)$ is the unitary operator determining the time evolution of the system. The formalism to describe the time evolution of the system and therefore the explicit expression of the unitary operators under consideration are resumed in Appendix \ref{time:evolution}. By making use of the expansion of the unitary operator in Eq.\eqref{unit.op.:expansion}, we can estimate the transition probability at the lowest order: 
\begin{align}
P(t)=\left(\frac{\epsilon}{\hbar}\right)^2\left\lvert\langle \Psi_f\lvert\hat U_0(t) \hat{U}_{\textrm{dr}}(t)\int_0^t dt'\hat{\tilde{H}}_{\textrm{I}}(t') \rvert \Psi_i\rangle\right\rvert^2,
\label{P:1}
\end{align}
where we collected the initial and final states as follows: $\lvert \Psi_i\rangle=\rvert \psi_i; 0_k,\bar 0_{k'}\rangle$ and $\lvert \Psi_f\rangle=\rvert \psi_f; 1_k,\bar 1_{k'}\rangle$.
Inserting the effective interaction Hamiltonian Eq.\eqref{app:tools} into Eq.\eqref{P:1}, we obtain the explicit form of $P^{(1)}(t)$:
\begin{widetext}
\begin{align}
P(t)=&4\epsilon^2(\omega_k-\omega_{k'})^2\left\lvert\chi_1(t) t\sinc\left[\frac{(\omega_k+\omega_{k'}+\Omega)t}{2}\right]e^{\frac{i(\omega_k+\omega_{k'}+\Omega)t}{2}}\right.\nonumber\\
&\left.+\chi_2(t) t\sinc\left[\frac{(\omega_k+\omega_{k'}-\Omega)t}{2}\right]e^{\frac{i(\omega_k+\omega_{k'}-\Omega)t}{2}}+\chi_3(t)\int_0^t\xi(t')e^{i(\omega_k+\omega_{k'})t'}\right\rvert^2,
\label{P:1:gen:tot}
\end{align}
\end{widetext}
where we defined the functions
\begin{align}
\chi_1(t)=&\langle\psi_f\lvert e^{-i\Omega t\hat b^\dag\hat b}\hat{U}_{\textrm{dr}}(t)\hat b^\dag\rvert\psi_i\rangle\nonumber\\
\chi_2(t)=&\langle\psi_f\lvert e^{-i\Omega t\hat b^\dag\hat b}\hat{U}_{\textrm{dr}}(t)\hat b\rvert\psi_i\rangle\nonumber\\
\chi_3(t)=&\langle\psi_f\lvert e^{-i\Omega t\hat b^\dag\hat b}\hat{U}_{\textrm{dr}}(t)\rvert\psi_i\rangle.
\label{chi}
\end{align}
It is therefore clear that our model is able to describe the transition probability caused by both the action of the external drive (whose intensity and trajectory can be controlled manually \cite{fosco_dynamical_2022}) and the spontaneous conversion of phonons into fermions. Although the latest occurs only if the system already contains (bosonic) excitations, and hence it is not strictly ascribable to the stimulation of the quantum vacuum, henceforth in this manuscript we will conveniently call dynamical Casimir effect (DCE) any effect of fermion pair generation.

Apart from its causes, in order for the DCE to occur, we need to impose the resonance condition between the frequency of the harmonic oscillator and the frequencies of the fermion modes $k$ and $k'$, namely $\Omega=\omega_k+\omega_{k'}$. For large $t$, we can avoid secularities and drastically simplify Eq.\eqref{P:1:gen} as follows:
\begin{align}
P(t)\simeq&4\epsilon^2(\omega_k-\omega_{k'})^2\left\lvert \chi_2(t) t+\chi_3(t)\int_0^t\xi(t')e^{i\Omega t'}\right\rvert^2.
\label{P:1:gen}
\end{align}
We notice that both terms within modulus depend on the action of the external drive. However, while the auxiliar function $\chi_2(t)$ vanishes in the absence of phonons, the last term proportional to $\chi_3(t)$ vanishes by switching off the external drive. In our following analysis we will explore some concrete scenarios, wherein we account for different input states and the action of an impulsive external drive.
\subsection{Phononic Fock state}
Let us consider the transition probability between two Fock states, $\lvert\psi_i\rangle\equiv\lvert j\rangle$ and $\lvert\psi_f\rangle\equiv\lvert l\rangle$.
In this case, it is possible to show that the auxiliar functions in Eq.\eqref{chi} are proportional to the matrix elements of the time dependent displacement operator $\chi_1(t)= \sqrt{j+1}e^{-il\omega t}[\hat D(\Lambda_t)]_{lj+1}$,  $\chi_2(t)= \sqrt{j}e^{-il\omega t}[\hat D(\Lambda_t)]_{lj-1}$ and $\chi_3(t)= e^{-il\omega t}[\hat D(\Lambda_t)]_{lj}$, where $\Lambda_t\equiv\Lambda(t)$ and the displacement operator is defined as usual $\hat D(\lambda)\equiv e^{\lambda \hat b^\dag-\lambda^*\hat b}$. We remind that the matrix elements of the displacement operator in the base of the Fock states are well-known in literature \cite{cahill_ordered_1969}:
\begin{align}
[\hat D(\lambda)]_{lj}=\sqrt{\frac{j!}{l!}}\lambda^{l-j}e^{-\lvert\lambda\rvert^2/2}L_j^{(l-j)}(\lvert\lambda\rvert^2),
\end{align}
with $l\ge j$, and 
\begin{align}
[\hat D(\lambda)]_{lj}=\sqrt{\frac{l!}{j!}}(-\lambda^*)^{j-l}e^{-\lvert\lambda\rvert^2/2}L_l^{(j-l)}(\lvert\lambda\rvert^2),
\end{align}
with $j\ge l$, where $L_n^{k}(x)$ are Laguerre polynomials.

By substituting the auxiliary functions into Eq.\eqref{P:1:gen} we obtain the transition probability:
\begin{align}
P(t)\simeq&4\epsilon^2(\omega_k-\omega_{k'})^2\left\lvert t\sqrt{j}[\hat D(\Lambda_t)]_{lj-1} \right.\nonumber\\
&\left.+[\hat D(\Lambda_t)]_{lj}\int_0^t\xi(t')e^{i(\omega_k+\omega_{k'})t'}\right\rvert^2.
\label{P:1:res:fock}
\end{align}
We notice that this formula is drastically simplified by switching off the external drive, providing a direct relation between the transition probability and the initial number of mechanical excitations:
\begin{align}
P(t)\simeq&4j\,\epsilon^2(\omega_k-\omega_{k'})^2t^2.
\label{P:1:nodrive}
\end{align}
On the other hand, the presence of the drive ensures a non-vanishing transition probability even in the phononic vacuum, i.e.  $\lvert\psi_i\rangle=\lvert\psi_f\rangle=\lvert 0\rangle$, which reduces Eq.\eqref{P:1:res:fock} as follows:
\begin{align}
P(t)\simeq&4\epsilon^2(\omega_k-\omega_{k'})^2e^{-\lvert\Lambda_t\rvert^2}\left\lvert \int_0^t\xi(t')e^{i(\omega_k+\omega_{k'})t'}\right\rvert^2.
\label{P:1:vacuum}
\end{align}
This equation describes the probability to excite a fermion pair by directly converting all phonons produced by the external drive, namely without enabling the excitation of the vibrational mode.
\subsection{Phononic squeezed-coherent state}
As a second example, we want to estimate the transition probability to generate fermion pairs when an initial phononic squeezed-coherent state is fully converted in a vacuum state, namely when
$\lvert\psi_f\rangle=\lvert 0\rangle$ and $\lvert\psi_i\rangle\equiv\hat D(\beta)\hat S(\zeta)\lvert 0\rangle$, with squeezing operator $\hat S(\zeta)\equiv e^{\frac{1}{2}(\zeta^* \hat b^2-\zeta\hat b^{\dag 2})}$ having squeezing parameter $\zeta=r e^{i\phi}$, whereas the coherent paremeter is $\beta=\lvert\beta\rvert e^{i\theta}$. 
For this transition probability, the $\chi_i$ functions in Eq.\eqref{chi} assume a more complicated form with respect to the case studied above; indeed, they read
\begin{align}
\chi_1(t)=&-\Lambda_t^*c_0(\beta+\Lambda_t,\zeta) e^{\frac{1}{2}(\Lambda_t \beta^*-\Lambda_t^*\beta)},\nonumber\\
\chi_2(t)=&[c_1(\beta+\Lambda_t,\zeta)-\Lambda_t c_0(\beta+\Lambda_t,\zeta)]e^{\frac{1}{2}(\Lambda_t \beta^*-\Lambda_t^*\beta)},\nonumber\\
\chi_3(t)=&c_0(\beta+\Lambda_t,\zeta) e^{\frac{1}{2}(\Lambda_t \beta^*-\Lambda_t^*\beta)},
\end{align}
where $c_n(\beta+\Lambda_t,\zeta)$ are the coefficients of the squeezed-coherent state in the Fock basis with $\beta+\Lambda_t$ coherent parameter and $\zeta$ squeezing parameter. Such coefficients are known in literature and will be not reported here; their explicit form was firstly calculated in \cite{gong_expansion_1990}.

We conclude this section by reporting the transition probability when the external drive is switched off. In case of resonance, this reads:
\begin{align}
P(t)=&4\epsilon^2(\omega_k-\omega_{k'})^2t^2\lvert c_1(\beta,\zeta)\rvert^2\nonumber\\
=&4\epsilon^2\lvert\gamma\rvert^2(\omega_k-\omega_{k'})^2t^2 \frac{e^{-\lvert\beta\rvert^2(1+(\cos(2\theta-\phi)\tanh (r))}}{\cosh^{3}(r)},
\label{P:1:gen:sc}
\end{align}
with $\gamma=\beta\cosh(r)+\beta^*e^{i\phi}\sinh(r)$.

\begin{figure*}
	\centering
	\includegraphics[width=1\linewidth]{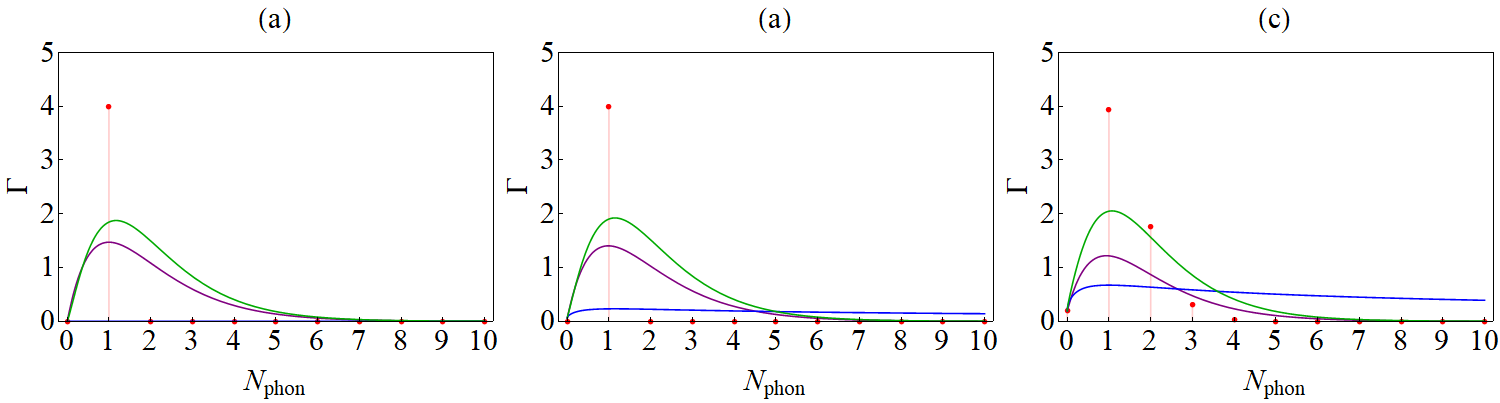}
	\caption{Comparison between the transition probabilities $\Gamma_{\textrm{F}}(j,g)$ (red, dotted), $\Gamma_{\textrm{C}}(\beta,g)$ (purple), $\Gamma_{\textrm{S}}(\zeta,g)$ (blue), $\Gamma_{\textrm{SC}}(\beta,\zeta,g)$ (green) and different drive strengths: (a) $g=0$, (b) $g=0.5$ and (c) $g=1$.}
	\label{gamma}
\end{figure*}

\subsection{Impulsive drive}
In order to concretely analyse and compare the trend of the transition probabilities studied so far, in this section we make use of a specific external drive.
For the sake of simplicity, we will always assume that the transition occurs towards the bosonic vacuum as output state, $\lvert\psi_f\rangle=\lvert 0\rangle$.
 As first case, we start from the transition probability between a bosonic Fock state and the vacuum, which becomes
 \begin{align}
P(t)\simeq&4\frac{\epsilon^2}{j!}\lvert\Lambda_t\rvert^{2j-2}(\omega_k-\omega_{k'})^2 e^{-\lvert\Lambda_t\rvert^2}\nonumber\\
&\times\left\lvert j t -\Lambda_t^*\int_0^t\xi(t')e^{i(\omega_k+\omega_{k'})t'}\right\rvert^2.
\label{P:1:res:fock1}
\end{align}
We now account for a specific complex external drive, whose real and imaginary parts respectively read:

\begin{align}
\lambda_{xb}(t)=&-\frac{g\nu}{2}e^{-\nu t}\cos(\Omega t),\nonumber\\
\lambda_{pb}(t)=&-\frac{g\nu}{2}e^{-\nu t}\sin(\Omega t),
\label{exdr}
\end{align}
where $g$ is the drive strength and $\nu$ is a parameter having the dimension of a frequency. When $\nu\gg\Omega$, such drive represents a short pulse, which kicks the wall and triggers the classical harmonic motion; indeed, after a short transient, $t\gg 1/\nu$, the kinematics of the wall is perfectly described by the classical trajectory $x(t)=L_0(1+\epsilon g\sin(\Omega t))$ \cite{ferreri2022interplay}. 
Clearly, the drive can always be switched off by imposing $g=0$.

Always assuming the resonance condition $\Omega=\omega_k+\omega_{k'}$, we are able to write a more convenient form of the transition probability by firstly substituting Eq.\eqref{exdr} into Eq.\eqref{P:1:res:fock1}, and then taking the limit of large time in order to avoid secularities. This reduces Eq.\eqref{P:1:res:fock1} to:

\begin{align}
P_\textrm{F}(t)\approx&\epsilon^2(\omega_k-\omega_{k'})^2t^2 \left(\frac{g}{2}\right)^{2j-2} e^{- g^2/4}\left(\frac{g^2+8j}{8\sqrt{j!}}\right)^2.
\label{P:1:res:fock2}
\end{align}

We can now apply the same procedure to the initial phononic squeezed-coherent state. However, since the formula of the transition probability is cumbersome, we report it in Appendix \ref{app2}, while here we restrict ourselves to give the transition probabilities from a coherent state and a squeezed state, respectively:
\begin{subequations}\label{S&C}
\begin{align}
P_\textrm{C}(t)\approx&\epsilon^2(\omega_k-\omega_{k'})^2t^2 e^{- g^2/4-\lvert\beta\rvert^2+g\beta\sin(\theta)}\nonumber\\
&\times\left(\frac{g^2}{4}+4\lvert\beta\rvert^2+2g\beta\sin(\theta)\right),
\label{P:1:res:coh}\\
P_\textrm{S}(t)\approx&\epsilon^2g^2(\omega_k-\omega_{k'})^2t^2 \frac{e^{- \frac{g^2}{4}(1-\cos(\phi)\tanh(r))}}{8\cosh^3(r)}\nonumber\\
&\times\left(5\cosh(2r)+4\sinh(2r)\cos(\phi)-3\right).
\label{P:1:res:squ}
\end{align}
\end{subequations}
Such formulas are easily obtained from Eq.\eqref{Prob:SC} by setting $r=0$ and $\beta=0$, respectively.

\subsection{Theoretical results}

Both Eqs.\eqref{P:1:res:fock2} and \eqref{S&C} describe the transition probability to generate the fermion pair both under the action of the external drive in Eq.\eqref{exdr} and by completely depleting the vibrational mode.
In order to compare these formulas, we focus our attention on the auxiliary functions:
\begin{subequations}\label{gammaFSC}
\begin{align}
\Gamma_{\textrm{F}}(j,g)=& \left(\frac{g}{2}\right)^{2j-2} e^{-\lvert g\rvert^2/4}\left(\frac{g^2+8j}{8\sqrt{j!}}\right)^2,\label{gammaf}\\
\Gamma_{\textrm{C}}(\beta,g)=&e^{- g^2/4-\lvert\beta\rvert^2+g\beta\sin(\theta)}\nonumber\\
&\times\left(\frac{g^2}{4}+4\lvert\beta\rvert^2+2g\beta\sin(\theta)\right),
\label{gammac}\\
\Gamma_{\textrm{S}}(\zeta,g)=&g^2 \frac{e^{- \frac{g^2}{4}(1-\cos(\phi)\tanh(r))}}{8\cosh^3(r)}\nonumber\\
&\times\left(5\cosh(2r)+4\sinh(2r)\cos(\phi)-3\right)\label{gammas},
\end{align}
\end{subequations}
and $\Gamma_{\textrm{SC}}(\beta,\zeta,g)$ (reported in the Appendix \ref{app2}, Eq.\eqref{gammasc}), so that we can write Eq.\eqref{P:1:res:fock2}, Eqs.\eqref{S&C} and Eq.\eqref{Prob:SC} in a more compact form:
\begin{subequations}\label{probFSC}
\begin{align}
P_\textrm{F}(t)=&\epsilon^2(\omega_k-\omega_{k'})^2t^2\Gamma_{\textrm{F}}(j,g),\\
P_\textrm{C}(t)=&\epsilon^2(\omega_k-\omega_{k'})^2t^2\Gamma_{\textrm{C}}(\beta,g)\\
P_\textrm{S}(t)=&\epsilon^2(\omega_k-\omega_{k'})^2t^2\Gamma_{\textrm{S}}(\zeta,g)\\
P_\textrm{SC}(t)=&\epsilon^2(\omega_k-\omega_{k'})^2t^2\Gamma_{\textrm{SC}}(\beta,\zeta,g).
\end{align} 
\end{subequations}
It becomes clear that Eq.\eqref{gammaFSC} and Eq.\eqref{gammasc} fully identify the transition probabilities of interest, and we can therefore focus our attention on them rather than on Eq.\eqref{probFSC}. Hence, in Fig.\ref{gamma} we plot the auxiliary functions at different drive strengths. Such graphs are realized by exploiting the fact that, as long as $\beta$ and $\zeta$ are real numbers, the average phonon number $N_\textrm{phon}=\langle\hat b^\dag\hat b\rangle$ depends on $j$, $\beta$ or $\zeta$ in an univocal way, and therefore we can invert them in order to use $N_\textrm{phon}$ as unique variable. This means that we can plot the functions  $\Gamma_{\textrm{F}}(j=N_\textrm{phon},g)$, $\Gamma_{\textrm{C}}(\beta=\sqrt{N_\textrm{phon}},g)$, $\Gamma_{\textrm{S}}(\zeta=\arsinh(\sqrt{N_\textrm{phon}}),g)$ and $\Gamma_{\textrm{SC}}(\beta=\sqrt{N_\textrm{phon}/2},\zeta=\arsinh(\sqrt{N_\textrm{phon}/2}),g)$. In the latest case we supposed that the displacement and the squeezing operations contribute to the phonon number equally, i.e. $\lvert\beta\rvert^2=\sinh^2 r=N_\textrm{phon}/2$.

Our results in Fig.\ref{gamma}a clearly show that, as long as the external drive is switched off, the transition probabilities reach their maximum at $N_\textrm{phon}=1$, except for $\Gamma_{\textrm{SC}}(\beta,\zeta,g)$, which is slightly shifted due to the squeezing contribution. This behaviour comes not as a surprise, since in the interaction Hamiltonian the generation of two fermions occurs at the price of only one phonon.
The fact that the blue line, namely the function $\Gamma_{\textrm{S}}(\zeta,g)$, is zero everywhere, stems from the boson statistics of the squeezing state, which is zero whenever the number of phonons is odd \cite{barnett2002methods}. Hence, supposing that the initial state contains two phonons, only one converts into the fermion pair, and consequently the system never reach the state $\lvert\psi_f\rangle=\lvert 0\rangle$.

By turning the external drive on, the contribution of the vacuum state to the transition probability becomes appreciable. Such contribution was analytically investigated and reported in Eq.\eqref{P:1:vacuum}, and describes the probability to directly convert the mechanical motion of the wall, caused by the external drive, into fermion pairs. This process can be interpreted as if the drive supplies the system with additional phonons. This interpretation would also explain the slight shift leftwards of the peak in the purple and green line; in fact, if the drive supplied additional phonons, we would need a reduced amount of initial coherent phonons $\lvert\beta\rvert^2$ in order to generate the same number of fermions. Finally, we notice that the presence of the drive, acting as a displacement, also enables the total conversion of phonons at higher Fock number and even when phonons are initially prepared in the squeezed vacuum.

\section{Extension to multiple bags}\label{extension}
The potential in Eq.\eqref{pot} embodies the boundary conditions for a spinor field confined in a cavity; indeed, the two delta functions constitute the walls of the cavity. In this section, we want to extend the analysis to a field confined in a region containing N equally-spaced spikes, some of them undergoing a position fluctuation $dL_q$ \cite{doi:10.1063/1.5106409}. The generalization of the potential becomes: 
\begin{align}
V(x)=2\sum_{q=0}^{N-1}\delta(x-q L-l_q dL_q).
\end{align}
where the parameters $l_q=0,1$ determine which bag undergoes the fluctuation. 

Assuming that all spikes have the same mass, the generalization of the quantization protocol reported in Section \ref{Hamilton} is straightforward. Once we obtain the extended classical Hamiltonian, we promote the increments $dL_q$ to quantum operators as follows
$\delta L_q\rightarrow\delta L_0 (\hat{b}_q^\dag+\hat{b}_q)$.
This step expands the number of bosonic degrees of freedom up to $N_l\le N$, where $N_l$ is the number of fluctuating walls. The total Hamiltonian therefore reads:
\begin{align}
H(t)=&H_0+\epsilon H_I+\hat H_{\textrm{dr}}(t),\\
H_0=&\sum_n\omega_n(\hat c_n^\dag\hat c_n+\hat d_n^\dag\hat d_n)+\sum_{l}^{N_l}\Omega_l\hat b_l^\dag\hat b_l,\\
\hat H_I=&-2\sum_{l}^{N_l}\sum_{n m}(-1)^{n+m}\left[(\omega_n+\omega_m)(\hat c_n^\dag \hat c_m+\hat d_m^\dag\hat d_n)\right.\nonumber\\
&\left.+(\omega_n-\omega_m)(\hat d_n\hat c_m+\hat d_m^\dag\hat c_n^\dag)\right](\hat b_l^\dag+\hat b_l),\label{HI2}\\
\hat H_{\textrm{dr}}(t)=&\sum_{l}^{N_l}\left(\lambda_l(t)\hat b_l^\dag+\lambda_l^*(t)\hat b_l\right),
\end{align}
where $\Omega_l$ are the oscillation frequencies of the walls, and we introduced the set of external drives $\lambda_l(t)$. 

As seen in Section \ref{corr}, the presence of counterrotating terms in the interaction Hamiltonian alters the ground state of the total system, which does not correspond to the vacuum state anymore. Since the harmonic oscillators do not interact with each other at the lowest order, the correction to the ground state and the vacuum energy can be easily calculated:
\begin{align}
\lvert\Psi_{0}\rangle=&\lvert 0; 0,\bar 0\rangle\nonumber\\
&+2\epsilon\sum_{l}^{N_l}\sum_{nm}\frac{(-1)^{n+m}(\omega_n-\omega_m)}{\omega_n+\omega_m+\Omega_l}\lvert 1_l; 1_n,\bar 1_{m}\rangle,\\
    \Delta E=&-\sum_{l}^{N_l}\sum_{nm}\frac{4\epsilon^2(\omega_n-\omega_m)^2}{\omega_n+\omega_m+\Omega_l}.
\end{align}
\section{Conclusions}\label{concl}
In this manuscript we have presented a technique for the introduction of a vibrational mode in a cavity system confining a fermionic field. The new mode, having bosonic nature, stems from the fluctuation of the position of one cavity wall, and represents a further quantum degree of freedom of the system. Interestingly, our quantization protocol of the harmonic motion of the wall directly provides the coupling terms between the bosonic subsystem and the trapped field.

We have demonstrated that, due to the presence of counterrotating terms, the interaction between the spinor field and the oscillating wall brings about a modification of the ground state, which does not correspond to the vacuum state anymore, but is expressed as a superposition of terms containing virtual particles. We showed that such virtual particles induce a negative shift in the Casimir energy of the system.

The interaction Hamiltonian expresses the possibility to convert vibrational excitations of the wall into fermion pairs. Hence, we proved that the excitation transfer strictly depends on the initial state of the phononic mode. As a specific scenario, we investigated the excitation of two fermionic modes, generating a particle and an antiparticle, following the total depletion of the bosonic mode. We showed how the manipulation of an external drive acting on the bosonic mode permits to control the transition probability, and enables the conversion of a higher number of phonons into fermions.

Finally, we generalized our model by considering a set of oscillating walls. The new total system consists therefore of a field interacting with a set of independent quantum harmonic oscillators. Also in this case, the correction to both the ground state and the Casimir energy was calculated.

The techniques and the concepts presented in this work allow to explore the thin borders between cavity electrodynamics, optomechanics and quantum field theory.
\section{Acknowledgments}
The author thanks David Edward Bruschi for the helpful comments.
The author acknowledges the 1003
program Geqcos, sponsored by the German Federal Min- 1004
istry of Education and Research, under the funding program 1005
“quantum technologies–from basic research to the market”, 1006
No. 13N15685.

\bibliographystyle{apsrev4-2}
\bibliography{biblio}
\onecolumngrid
\appendix
\section{Time evolution}\label{time:evolution}
In this work we are interested in investigating the transition probability to convert vibrational excitations into fermion pairs. To do that, we need an agreeable expression for the time evolution operator $\hat{U}(t)$ \cite{Qvarfort_2019, Qvarfort_2020}. 
In presence of a time-dependent Hamiltonian $\hat{H}(t)$, this reads
\begin{align}\label{time:evolution:operator}
\hat{U}(t)=\overset{\leftarrow}{\mathcal{T}}\exp\left[-\frac{i}{\hbar}\int_0^{t}\,dt'\,\hat{H}(t')\right],
\end{align}
where $\overset{\leftarrow}{\mathcal{T}}$ stands for the time-ordering operator.

The total Hamiltonian in Eq.\eqref{htot} consists in three parts, one of which depends on time. Although they do not commute with each other, it turns out to be more convenient to rearrange the time evolution operator such that we can split the action of the single Hamiltonian contributions as follows 
\begin{align}\label{time:evolution:operator}
\hat{U}(t)=\hat{U}_0(t)\hat{U}_{\textrm{dr}}(t)\hat{U}_{\textrm{I}},
\end{align}
modulo an overall complex phase that has no physical significance. In Section \ref{TP} this choice allows us to address the analysis of the transition probability via perturbation theory in a more convenient way. In Eq.\eqref{time:evolution:operator} we distinguish the following unitary operators
\begin{align}\label{quantum:hamiltonian:terms}
\hat{U}_0(t):=&\exp\left[-i\,\hat{H}_0 t\right],\\ 
\hat{U}_{\textrm{dr}}(t):=&\overset{\leftarrow}{\mathcal{T}}e^{-i\int_0^t dt'\hat{\tilde{H}}_{\textrm{dr}}(t')}\simeq\exp\left\{\,\Lambda^*(t)\hat{b}^\dag-\Lambda(t)\hat{b}\right\},\label{ED}
\\ 
\hat{U}_{\textrm{I}}(t):=&\overset{\leftarrow}{\mathcal{T}}e^{-i\epsilon\int_0^t dt'\hat{\tilde{H}}_{\textrm{I}}(t')}\label{UI},
\end{align}
where we defined the effective drive, $\hat{\tilde{H}}_{\textrm{dr}}(t):=\hat{U}^\dag_0(t)\hat{H}_{\textrm{dr}}(t)\hat{U}_0(t)$, and the time-dependent displacement parameter
\begin{align}
\Lambda(t):=&\int_0^tdt'\left[\lambda_{x}(t')\sin(\Omega t')-\lambda_{p}(t')\cos(\Omega t')+i\left(\lambda_{x}(t')\cos(\Omega t')+\lambda_{p}(t')\sin(\Omega t')\right)\right].
\end{align}
The approximation in Eq.\eqref{ED} is justified by the presence of a global phase, which does not have any physical consequence on our calculations and can therefore be omitted.

The unitary operator in Eq.\eqref{UI} contains the effective interaction Hamiltonian 
\begin{equation}
\hat{\tilde{H}}_{\textrm{I}}(t):=\hat{U}^\dag_{\textrm{dr}}(t)\hat{U}^\dag_0(t)\hat{H}_{\textrm{I}}(t)\hat{U}_0(t)\hat{U}_{\textrm{dr}}(t),
\label{H:tilde}
\end{equation}
whose explicit expression is 
\begin{align}
\hat{\tilde H}_I(t)=&-2\sum_{n m}(-1)^{n+m}\left[(\omega_n+\omega_m)(e^{i(\omega_n-\omega_m) t}\hat c_n^\dag \hat c_m-e^{-i(\omega_n-\omega_m) t}\hat d_n\hat d_m^\dag)\right.\nonumber\\
&\left.-(\omega_n-\omega_m)(e^{i(\omega_n+\omega_m) t}\hat c_n^\dag\hat d_m^\dag-e^{-i(\omega_n+\omega_m) t}\hat d_n\hat c_m)\right](e^{i\Omega t}\hat b^\dag+e^{-i\Omega t}\hat b+2\xi(t))
\label{app:tools}
\end{align}
where $\xi(t)=\Re\{\Lambda(t)\}\cos(\Omega t)-\Im\{\Lambda(t)\}\sin(\Omega t)$.
Such unitary operator depends on the adimensional amplitude $\epsilon$, which is supposed to be extremely small. This enables the expansion of Eq.\eqref{UI} at the lowest perturbative orders:
\begin{align}
U_I(t)\simeq& 1-i\epsilon\int_0^t dt'\hat{\tilde{H}}_{\textrm{I}}(t')-\epsilon^2\int_0^t dt'\hat{\tilde{H}}_{\textrm{I}}(t')\int_0^{t'} dt''\hat{\tilde{H}}_{\textrm{I}}(t'').
\label{unit.op.:expansion}
\end{align}
\section{Transition probability $P_{SC}(t)$}\label{app2}
The probability to convert all phonons prepared in a initial phononic squeezed-coherent state into a fermionic pair is given by:
\begin{align}
P_{SC}(t)\approx&\epsilon^2(\omega_k-\omega_{k'})^2t^2\exp\left\{-\frac{g^2}{4}-\lvert\beta\rvert^2+g\lvert\beta\rvert\sin(\theta)+\frac{\tanh(r)}{4}\left(g^2\cos(\phi)-4\lvert\beta\rvert^2\cos(2\theta-\phi)-4g\lvert\beta\rvert\sin(\theta-\phi)\right)\right\}\nonumber\\
&\times\sech(r)\bigg\{\sech(r)^2\left[3g\lvert\beta\rvert\sin(\theta)-\frac{3}{8}g^2+\cosh(2r)\left(\frac{5}{8}g^2+4\lvert\beta\rvert^2-g\lvert\beta\rvert\sin(\theta)\right)\right]\nonumber\\
&+[8\lvert\beta\rvert^2\cos(2\theta-\phi)+g^2\cos(\theta)+2g\lvert\beta\rvert\sin(\theta-\phi)]\tanh(r)\bigg\}.
\label{Prob:SC}
\end{align}
From this transition probability we define the auxiliary function:
\begin{align}
\Gamma_{\textrm{SC}}(\beta,\zeta,g)=&\exp\left\{-\frac{g^2}{4}-\lvert\beta\rvert^2+g\lvert\beta\rvert\sin(\theta)+\frac{\tanh(r)}{4}\left(g^2\cos(\phi)-4\lvert\beta\rvert^2\cos(2\theta-\phi)-4g\lvert\beta\rvert\sin(\theta-\phi)\right)\right\}\nonumber\\
&\times\sech(r)\bigg\{\sech(r)^2\left[3g\lvert\beta\rvert\sin(\theta)-\frac{3}{8}g^2+\cosh(2r)\left(\frac{5}{8}g^2+4\lvert\beta\rvert^2-g\lvert\beta\rvert\sin(\theta)\right)\right]\nonumber\\
&+[8\lvert\beta\rvert^2\cos(2\theta-\phi)+g^2\cos(\theta)+2g\lvert\beta\rvert\sin(\theta-\phi)]\tanh(r)\bigg\}.
\label{gammasc}
\end{align}

\end{document}